\documentclass[12pt]{article}
\renewcommand{\baselinestretch}{1.0}
\usepackage{amsmath,amssymb,amsthm,graphicx,latexsym}

\newcommand{\ed}{\end{document}}
\newcommand{\beq}{\begin{equation}}
\newcommand{\eeq}{\end{equation}}
\newcommand{\beqa}{\begin{eqnarray}}
\newcommand{\eeqa}{\end{eqnarray}}
\newcommand{\bc}{\begin{center}}
\newcommand{\ec}{\end{center}}

\newcommand{\ba}{\begin{array}}
\newcommand{\ea}{\end{array}}
\newcommand{\pa}{\partial}

\newcommand{\singlespacing}{\let\CS=\@currsize\renewcommand{\baselinestretch}{
1.0}\tiny\CS}
\newcommand{\doublespacing}{\let\CS=\@currsize\renewcommand{\baselinestretch}{
1.5}\tiny\CS}

\def\({\left(}
\def\){\right)}

\begin{document}
\begin{center}
\Large{\bf{Noncommutative Extension of AdS-CFT and Holographic Superconductors}}
\end{center}
\vspace{0.0cm}
\begin{center}
Souvik Pramanik \footnote{E-mail: souvick.in@gmail.com},
Sudipta Das \footnote{E-mail: sudipta$_-$jumaths@yahoo.co.in},
Subir Ghosh \footnote{E-mail: subir$_-$ghosh2@rediffmail.com}\\
\vspace{2.0 mm}
\small{\emph{Physics and Applied Mathematics Unit, Indian Statistical
Institute\\
203 B. T. Road, Kolkata 700108, India}} \\
\end{center}
\vspace{0.5cm}
\begin{abstract}
In this Letter, we consider a Non-Commutative (NC) extension of AdS-CFT correspondence and its effects on holographic superconductors. NC corrections are incorporated via the NC generalization of Schwarzschild black hole metric in AdS with the probe limit. We study NC effects on the relations connecting the charge density and the critical temperature of the Holographic Superconductors. Furthermore, condensation operator of the superconductor has been analyzed. Our results suggest that generically, NC effects increase the critical temperature of the holographic superconductor.
\end{abstract}
\vskip .1cm

\section{Introduction}
In recent years AdS/CFT correspondence, proposed by Maldacena \cite{adscft}, has captured the attention of both High Energy and Condensed Matter theorists, since it can address issues in strongly interacting systems in the latter one (that are otherwise intractable in conventional Condensed Matter framework), by exploiting results obtained in weakly coupled systems in the former. In particular, there exists explicit mapping between relevant operators and parameters of a field theory in the bulk AdS space-time to those of a Conformal Field Theory living in the (one dimension lowered) boundary. It was  shown by Gubser \cite{gubser} that a simple theory of Abelian Higgs model in AdS space can lead to a spontaneous symmetry breaking thereby inducing scalar hair near the black hole horizon. The AdS-CFT correspondence and its associated dictionary can lead to interesting analogies with thin-film superconductors. A still simpler variant that captures the essential physics of holographic superconductors was considered by  Hartnoll, Herzog and  Horowitz \cite{hart}, who took the so called probe limit where the Maxwell field and the scalar field do not generate back reactions on the metric. Operationally, this means that one can consider the effect of Schwarzschild-AdS metric on scalar and Maxwell fields, instead of taking the background metric to be Reissner-Nordstrom in AdS. In fact, afterward the latter framework was studied in \cite{hart1} (for detailed review see \cite{holo}).

In this perspective, our aim is to study the effects of Non-Commutative (NC) geometry on AdS-CFT correspondence and subsequently on the properties of holographic superconductors. In the present work, we focus on the probe limit and will pursue the full theory including back reactions in a later publication.

Noncommutativity in spacetime was introduced long ago by Snyder \cite{sny} in the hope of removing short distance singularities in quantum field theory, but it was not successful. Later on, NC field theory was resurrected by Seiberg and Witten \cite{string}, who demonstrated that in the low energy limit open strings, attached to $D$-branes, induced noncommutativity in the $D$-branes. In \cite{string} rules were provided for extending QFTs to NC QFTs, where normal products between local fields were replaced by $*$-products so that NC QFTs can be studied perturbatively for small NC parameter $\theta$. Furthermore, NC gauge theories had to be treated in a special way by incorporating the Seiberg-Witten map \cite{string} (for a review see \cite{nft}). Recently Nicolini, Smailagic and Spalucci \cite{nicolini, nico1} have given a new NC extension of Schwarzschild metric by directly solving the Einstein's equation with a smeared matter source, which has the form of a gaussian distribution that incorporate the NC effect as a minimum width $\theta$. The black hole singularity was successfully removed in this scenario. In a sense, the original motivation of Snyder \cite{sny} was partly fulfilled albeit in a classical context. NC effects on salient features of black hole, such as Hawking radiation, have been studied using this $\theta$-corrected metric \cite{banerjee}. In these analysis $\theta$ is the NC parameter. Hawking-Page crossover with such NC black hole metric in AdS has been studied in \cite{nico2}. Recently, effects of noncommutativity on thermalization processes for the NC black hole backgrounds have been studied in \cite{zeng}.

In the present paper, we aim to study the bulk NC effect on holographic superconductors in the probe limit approximation. It is perhaps acceptable that (at least to the lowest non-trivial order of $\theta$) NC effect does not change the asymptotic behaviors of bulk fields qualitatively. This means that the functional forms remain unchanged whereas the numerical parameters undergo the NC corrections. This allows us to use the same (canonical) AdS-CFT dictionary in order to compute the $\theta$-corrected relation between the critical temperature and the charge density of the holographic superconductor and thereafter the condensate-temperature relation. As we have mentioned our results are valid in the probe limit domain.

The paper is organized as follows: In Section 2, we introduce the NC-AdS black hole metric and define the action for an Abelian gauge field (coupled with a scalar) in this NC spacetime background. In Section 3, we study the asymptotic behavior of the gauge and scalar fields. Depending on these we proceed to analyze the relation between the critical temperature and the charge density in Section 4. Afterward, in Section 5 we compute the critical exponents and condensation values. Finally we conclude with a discussion of our findings in Section 6.

\section{Noncommutative black hole in $AdS_4$ and equation of motions}

We start with the gravitational action in noncommutative AdS background, where gravity is coupled to a $U(1)$ gauged  charged scalar field $\psi $, given by
\beq
S=\int d^4 x \sqrt{-g} {\mathcal{L}} = \int d^4 x \sqrt{-g} \left[\frac{1}{16 \pi G} \left( R+\frac{6}{L^2}\right)-\frac{1}{4}{F}_{\mu\nu}
{F}^{\mu\nu}-|\partial_\mu \psi-i q{A}_\mu \psi|^2-m^2|\psi|^2\right]. \label{action}
\eeq
In the above $g_{\mu \nu}$ is the metric tensor, $L$ is the AdS radius, ${F}_{\mu\nu}$ is the Maxwell field and $\psi$ is the scaler field of Higgs.  Throughout this paper, we work in the system of natural units with $c=\hbar=k_B=1$. As we have already discussed previously, the noncommutativity emerges here from both the matter and electromagnetic source terms, are smeared ones with gaussian features as considered in \cite{nicolini, nico1, nico3}. The solution of the above action is given by a NC charged $AdS_4$ black hole metric \cite{nicolini, nico1, nico3},
\beq
ds^2=-f_1(r)dt^2+\frac{dr^2}{f_1(r)}+r^2 d \Omega^2; $$$$
f_1(r)=K-\frac{4 M G}{r \sqrt{\pi}} \gamma(3/2, r^2/4 \theta)+\frac{G Q^2 }{\pi r^2}
\left[\gamma^2(1/2, r^2/4 \theta) - \frac{r}{\sqrt{2 \theta}} \gamma(1/2, r^2/2 \theta) + \sqrt{\frac{2}{\theta}} r \gamma(3/2, r^2/4 \theta)\right]+\frac{r^2}{L^2}, \label{metric}
\eeq
where $\gamma(s, x) = \int_0^x t^{s-1} e^{-t} dt$ is the lower incomplete Gamma function and $Q$ is the total charge of the black hole. In (\ref{metric}), $K$ represents the curvature which take the values $K=0,+1,-1$ corresponding to planar, spherical and hyperbolic spacetime respectively. For $\theta \rightarrow 0$ the above metric reduces to the usual Reissner-Nordstrom $AdS_4$ form. Since we are interested on asymptotic behavior of AdS, and as stipulated earlier we restrict ourselves to the probe limit, the $Q^2$-dependent back reaction terms are not taken into account in $f_1(r)$. Therefore the metric (in probe limit) reduces to
\begin{eqnarray}
ds^2 &=& -f_1(r)dt^2+\frac{dr^2}{f_1(r)}+r^2 d\Omega^2,\nonumber\\
f_1(r) &=& K-\frac{4 M G}{r \sqrt{\pi}} \gamma(3/2, r^2/4 \theta)+\frac{r^2}{L^2}, \label{metric2}
\end{eqnarray}
The outer horizon radius $r_+$ for this black hole is obtained by solving
\beq K-\frac{4 M G}{r_+ \sqrt{\pi}} \gamma(3/2, r_+^2/4 \theta)+\frac{r_+^2}{L^2} =0. \label{bhhorizon} \eeq
In this work, we intend to study the large black holes, i.e. where $\frac{r_+^2}{4\theta}\gg 1$, and hence the lower incomplete gamma function in (\ref{bhhorizon}) can be approximated by the exponential form \cite{nicolini}. Moreover, as our goal in this paper is to study the properties of a holographic superconductor through AdS-CFT correspondence, at the AdS boundary (where $r$ is large and hence $\frac{r^2}{4\theta}\gg 1$, since $\theta$ is small), the exterior metric (\ref{metric2}) becomes
\begin{eqnarray}
ds^2 &=& -f_1(r)dt^2+\frac{dr^2}{f_1(r)}+r^2 ( dx^2 + dy^2 ),\nonumber\\
f_1(r) &=& K - \frac{2 M G}{r} + \frac{r^2}{L^2} + \frac{2 M G}{\sqrt{\pi \theta}}~ e^{-\frac{r^2}{4 \theta}} \label{metric-3}.
\end{eqnarray}
We consider gauge field $A_\mu$ to have only temporal component as it is customary \cite{gubser, holo}, i.e. $A_\mu=(\phi(r),0,0,0)$ and $\psi=\psi(r)$. With these assumptions, the action (\ref{action}) simplifies to
\begin{equation}
S = \int d^4x \sqrt{- g} \left[\frac{1}{16\pi G} \left(R+\frac{6}{L^2}\right)-\frac{1}{2}g^{rr}g^{tt} (\partial_r\phi)^2 - g^{rr}(\partial_r\psi)(\partial_r\psi^*) - g^{tt} \phi^2|\psi|^2 - m^2|\psi|^2\right]. \label{action1}
\end{equation}
We now derive the equations of motion for the scaler fields $\psi$ and the gauge potential $\phi$ from the action (\ref{action1}). The equation of motion for scalar field $\psi$ is given by
\beq
\psi''+\left(\frac{f_1'(r)}{f_1(r)}+\frac{2}{r}\right)\psi'-\frac{m^2}{f_1(r)}
\psi+\frac{ \phi^2}{f_1^2(r)}\psi = 0, \label{diffeqsi} \eeq
and the equation of motion for the gauge potential $\phi$ is given by
\beq
\phi'' + \frac{2}{r}\phi' - \frac{2|\psi|^2}{f_1(r)}\phi = 0. \label{diffeqfi}
\eeq
Equations (\ref{diffeqsi}) and (\ref{diffeqfi}) are the governing equations of our noncommutative model.

\section{Asymptotic behavior of $\psi$ and $\phi$}

Let us study the asymptotic behavior of $\psi$ and $\phi$, as these are relevant in the AdS-CFT correspondence. In this context, we take the metric (\ref{metric-3}) to be asymptotic one and following \cite{nico1, banerjee} we expand the subsequent terms throughout this work to $O(\frac{2 G M}{\sqrt{\pi\theta}}e^{-\frac{r^2}{4\theta}})$. Note that, though $1/\sqrt{\theta}$ is large (since $\theta$ is a non-zero small quantity and has an established upper bound \cite{jac}), in the asymptotic limit (for large $r$) the exponential damping term $e^{-\frac{r^2}{4\theta}}$ dominates over $1/\sqrt{\theta}$ and makes the overall quantity $(\frac{2 G M}{\sqrt{\pi\theta}}e^{-\frac{r^2}{4\theta}})$ small enough, so that the above expansion remains valid. With this argument we expand the governing equations for $\psi$ (\ref{diffeqsi}) and $\phi$ (\ref{diffeqfi}) to the first order of $O\left(\frac{2M G}{\sqrt{\pi\theta}}e^{-\frac{r^2}{4\theta}}\right)$. First, notice that with this expansion, the second term of (\ref{diffeqsi}) can be expanded like
$$
\left(\frac{f_1'(r)}{f_1(r)}+\frac{2}{r}\right)\psi= \left(\frac{f'(r)}{f(r)}+\frac{2}{r}\right)\psi
-\left(\frac{f'(r)}{f^2(r)}+\frac{r}{2\theta f(r)}\right)\frac{2M G}{\sqrt{\pi\theta}}
e^{-\frac{r^2}{4\theta}}\psi, $$
where $f(r) = K-\frac{2M G}{r}+\frac{r^2}{L^2}$. Performing similar kinds of expansion for rest of the terms in (\ref{diffeqsi}) and (\ref{diffeqfi}) the governing equation for $\psi$ and $\phi$ becomes
\beq
\psi''+\left(\frac{f'(r)}{f(r)}+\frac{2}{r}\right)\psi'-\frac{m^2}{f(r)}
\psi+\frac{ \phi^2}{f^2(r)}\psi -\left(\left(\frac{f'(r)}{f^2(r)}+\frac{r}{2\theta f(r)}\right)\psi'-\frac{m^2}{f^2(r)}\psi+\frac{2 \phi^2}{f^3(r)}
\psi \right)
\frac{2M G}{\sqrt{\pi\theta}}e^{-\frac{r^2}{4\theta}}=0 \label{siappro}\eeq
and
\beq \phi''+\frac{2}{r}\phi'-\frac{2 |\psi|^2}{f(r)}\phi
+  \frac{2 |\psi|^2}{f^2(r)} \times \frac{2M G}{\sqrt{\pi\theta}}e^{-\frac{r^2}{4\theta}} \phi = 0. \label{fiappro1}\eeq
As we are interested on studying the asymptotic behavior of the fields, we can consider a further approximation $\frac{1}{f(r)}\approx\frac{L^2}{r^2}$ (considering terms up to $\frac{1}{r^2}$ only), so that no higher order terms than $O(\theta /r^2)$ appear in the final equations;
{\it{asymptotically $r^2/ 4 \theta >> 1$ is equivalent to the limit where $\theta /r^2$ is very small}}.
With this  approximation, the above equation for $\psi$ can be
written (keeping terms only up to $\frac{1}{r^2}$) as
\beq \psi''+\frac{4}{r}\psi'+\frac{2}{r^2}\psi-\left(\frac{L^2}{2\theta
	r}\psi'\right)\frac{2M G}{\sqrt{\pi\theta}}e^{-\frac{r^2}{4\theta}}=0,
\label{siasymdiff}\eeq
where we have used the relation $m^2 L^2=-2$ \cite{gubser}. It is known that $\psi=\frac{C}{r}+\frac{D}{r^2}$ is a solution of the equation
$\psi''+\frac{4}{r}\psi'+\frac{2}{r^2}\psi=0.$ Therefore we assume that the solution of (\ref{siasymdiff}) to be of the form $\psi=\frac{C}{r}+\frac{D}{r^2}+\frac{2M G}{\sqrt{\pi\theta}}e^{-\frac{r^2}{4\theta}}\psi_1.$
Substituting this into (\ref{siasymdiff}) we see that the differential equation for $\psi_1$ becomes
\beq
\psi_1''+\left(\frac{4}{r}-\frac{r}{\theta}\right)\psi_1'+\left(\frac{r^2}{4\theta^2}-\frac{5}{2\theta}+\frac{2}{r^2}\right)\psi_1
=\frac{L^2}{2\theta r}\psi_0' = -\frac{L^2}{2\theta r}\left(\frac{C}{r^2}+\frac{D}{r^3}\right)\approx 0, \label{dif-si1} \eeq
where again we have considered terms up to $O(\frac{1}{r^2})$. The solution of (\ref{dif-si1}) is given by $\psi_1=\left(\frac{E}{r}+\frac{F}{r^2}\right)e^{\frac{r^2}{4\theta}}.$ Therefore the asymptotic behavior of $\psi$ can be expressed as
\begin{equation}
	\psi=\frac{C}{r}+\frac{D}{r^2}+\frac{2M G}{\sqrt{\pi\theta}}e^{-\frac{r^2}{4\theta}}\psi_1
	=\frac{C}{r}+\frac{D}{r^2}+\frac{2M G}{\sqrt{\pi\theta}}\left(\frac{E}{r}+\frac{F}
	{r^2}\right) = \frac{\psi^-}{r}+\frac{\psi^+}{r^2},\label{siasym}
\end{equation}
where $\psi^-=C+\frac{2M G}{\sqrt{\pi\theta}} E$ and $\psi^+=D+\frac{2M G}{\sqrt{\pi\theta}} F$. For later analysis we set $\psi^+ = 0$ and $\psi^- \simeq \langle J \rangle$ \cite{dibakar}. It is interesting to see from (\ref{siasym}) that the $\theta$-dependent part has the same structure as the usual one where  $\theta = 0$.

Using the same approximation $\frac{1}{f(r)}\approx \frac{L^2}{r^2}$ and substituting (\ref{siasym}) in (\ref{fiappro1}), while considering terms up to $\frac{1}{r^2}$, we finally get the equation for $\phi$ in the form
\beq
\phi''+\frac{2}{r}\phi' = 0,
\eeq
with the solution given by
\beq \phi=\mu-\frac{\rho}{r}.
\label{phiasyem1}\eeq
The constants $\mu$ and $\rho$ are interpreted respectively as chemical potential and charged density. From (\ref{phiasyem1}), it is observed that $\mu$ and $\rho$ are not affected by noncommutativity.

\section{Relation between critical temperature and charge density}
If we change the radial coordinate from $r$ to $z$ by the transformation $z=\frac{r_+}{r}$, then the above governing equations for $\psi$ (\ref{siappro}) and $\phi$ (\ref{fiappro1}) becomes:
\begin{eqnarray}
&&\psi''+\frac{f'(z)}{f(z)}\psi' -\frac{m^2 r_+^2}{z^4 f(z)} \psi+\frac{ r_{+}^2 \phi^2}{z^4 f^2(z)}\psi \nonumber\\
&& + \left(\left(-\frac{f'(z)}{f^2(z)}+\frac{r_+}{2\theta z^3 f(z)}\right) \psi' +\frac{m^2 r_+^2}{z^4 f^2(z)} \psi - \frac{2 r_+^2 \phi^2}{z^4 f^3(z)}\psi \right) \times \frac{2 M G}{\sqrt{\pi\theta}}e^{-\frac{r_+^2}{4\theta z^2}} =0 \nonumber\\ \label{psi-eqn-2}\\
&& \phi''-\frac{2 r_+^2|\psi|^2}{z^4 f(z)}\phi + \frac{2 r_+^2|\psi|^2}{z^4 f^2(z)} \frac{2M G}{\sqrt{\pi\theta}}e^{-\frac{r+^2}{4\theta z^2}}\phi = 0, \label{fi-eqn-2}
\end{eqnarray}
where the dashes represents the derivative with respect to $z$. In order to obtain the relation between the critical temperature $T_c$ and charge density $\rho$, we follow the technique exploited in \cite{dibakar}. At the critical temperature $T=T_c$, the scalar field $\psi$ vanishes, i.e. $\psi=0$, for which the equation (\ref{fi-eqn-2}) becomes
\beq
\phi'' = 0~~\Longrightarrow ~~\phi=a + b z. \label{fisol}
\eeq
By the transformation $z=\frac{r_+}{r}$ the solution region changes from $r_{+}\leq r <\infty$ to $1 \geq z > 0$. Since the horizon $f_1(r)=0$ is at $r=r_{+}$, by this transformation the horizon becomes at $z=1$ and asymptotic boundary becomes at $z=0$. At $T = T_c$, from the asymptotic solution of $\phi$ (\ref{phiasyem1}) we get the following relation: $\phi'(z)=-\frac{\rho}{r_{+c}}$ ($r_{+c}$ is the radius of the horizon at $T=T_c$). Comparing (\ref{fisol}) with this expression we have $b=-\frac{\rho}{r_{+c}}$.
Applying the horizon boundary condition $\phi(z=1)=0$ in (\ref{fisol}) we have $a = -b = \frac{\rho}{r_{+c}}$. Thus, the expression for the scalar potential $\phi$ at the critical temperature $T=T_c$ can be written as
\beq
\phi=\frac{\rho}{r_{+c}}(1-z)
\approx \lambda r_{+c}(1-z),~~~~~~~\lambda = \frac{\rho}{r_{+c}^2}.\label{fisol4}
\eeq
We are now going to investigate the boundary behavior of the scalar field $\psi$ as $T\rightarrow T_c$. Substituting the above form of the scaler potential $\phi$ (\ref{fisol4}) into (\ref{psi-eqn-2}) we have,
\begin{eqnarray}
	&&\psi''+\left[\frac{f'(z)}{f(z)}+\left(-\frac{f'(z)}{f(z)^2}+\frac{r_{+}^2}{2\theta z^3f(z)}\right)
	\frac{2M G}{\sqrt{\pi\theta}}e^{-\frac{r_{+}^2}{4\theta
			z^2}}\right]\psi'-\frac{r_{+}^2m^2}{z^4f(z)}
	\left[1-\frac{1}{f(z)} \frac{2M G}{\sqrt{\pi\theta}}e^{-\frac{r_{+}^2}{4\theta
			z^2}}\right]\psi \nonumber\\
	&& =-\lambda^2.\frac{r_{+}^4}{z^4f(z)^2}(1-z)^2 \left[1-\frac{2}{f(z)} \frac{2M G}{\sqrt{\pi\theta}}
	e^{-\frac{r_{+}^2}{4\theta z^2}}\right]\psi, \label{sidiffz}
\end{eqnarray}
where $f(z)= K-\frac{2M G}{r_{+}}z+\frac{r_{+}^2}{L^2 z^2}$. In order to study the behavior of $\psi$ near the asymptotic boundary ($z\rightarrow 0$) we can define (as $T\rightarrow T_c$) \cite{dibakar},
\beq
\psi(z)=\frac{\langle J\rangle}{\sqrt{2}r_{+}}z F(z), \label{siassumption}
\eeq
where $F(z)$ satisfies the boundary condition $F(0)=1$ and $F'(0)=0$. Substituting (\ref{siassumption}) in the above equation (\ref{sidiffz}) we have
\begin{eqnarray}
&& F''(z)+\left[\frac{2}{z}+\frac{f'(z)}{f(z)}+\left(-\frac{f'(z)}{f(z)^2}+\frac{r_
		{+}^2}{2\theta
		z^3f(z)}\right)\frac{2M G}{\sqrt{\pi\theta}}e^{-\frac{r_{+}^2}{4\theta
			z^2}}\right]F'(z) \nonumber\\
	&& +\left[\frac{f'(z)}{z f(z)}-\frac{r_{+}^2 m^2}{z^4f(z)}+\left(-\frac{f'(z)}{z f(z)^2}+\frac{r_{+}^2
		m^2}{z^4f(z)^2}+\frac{r_{+}^2}{2\theta z^4f(z)}\right)\frac{2M G}{\sqrt{\pi\theta}}
	e^{-\frac{r_{+}^2}{4\theta z^2}}\right]F(z) \nonumber\\
	&& +\lambda^2 \times \frac{r_{+}^4 (1-z)^2}{z^4f(z)^2}\left[1-\frac{2}{f(z)} \frac{2M G}{\sqrt{\pi\theta}}
	e^{-\frac{r_{+}^2}{4\theta z^2}}\right]F(z)=0. \label{Fdiff}
\end{eqnarray}
Now it is straightforward to cast (\ref{Fdiff}) into a Sturm-Liouville eigenvalue problem of the generic form \cite{dibakar}
\beq
T(z) F'(z) - Q(z)F(z) + \lambda^2 P(z) F(z) = 0 \label{eqn-lam-1}
\eeq
where
\begin{eqnarray}
T(z) &=& r_{+}L^2z^2f(z) \times e^{\frac{1}{f(z)}\times \frac{2M G}{\sqrt{\pi\theta}}e^{-\frac{r_{+}^2}{4\theta z^2}}}, \nonumber\\
Q(z) &=& -r_{+} L^2
\left(zf'(z)-\frac{m^2 r_+^2}{z^2}+\left(-\frac{zf'(z)}{f(z)}+\frac{m^2 r_+^2}{z^2 f(z)}+\frac{r_+^2}{2\theta z^2}\right)\frac{2M G}{\sqrt{\pi\theta}}e^{-\frac{r_{+}^2}{4\theta z^2}}\right)\times e^{\frac{1}{f(z)} \times \frac{2M G}{\sqrt{\pi\theta}}e^{-\frac{r_{+}^2}{4\theta z^2}}},\nonumber\\
P(z) &=& \frac{r_{+}^5 L^2}{z^2f(z)}(1-z)^2 \left(1-\frac{2}{f(z)} \times \frac{2M G}{\sqrt{\pi\theta}}
e^{-\frac{r_{+}^2}{4\theta z^2}}\right)e^{\frac{1}{f(z)} \times\frac{2M G}{\sqrt{\pi\theta}}e^{-\frac{r_{+}^2}{4\theta z^2}}} \label{tqp}
\end{eqnarray}
where $f(z)= K -\frac{2 M G}{r_{+}}z +\frac{r_{+}^2}{L^2 z^2}$. For a given choice of $F(z)$, the explicit form of $\lambda^2$ that can be obtained from the above expression (\ref{eqn-lam-1}) is given by
\beq
\lambda^2=\frac{\int_0^1~\left(T(z)\left\{F'(z)\right\}^2+Q(z)\left\{F(z)\right\}^2 \right) dz}
{\int_0^1~P(z)\left\{F(z)\right\}^2 dz}. \label{lambda-equ}
\eeq
The structure of $F(z)$ can be chosen to be $F(z)=1-c z^2$ \cite{hart, holo} which satisfies the boundary conditions $F(0)=1$ and $F'(0)=0$. Here, $c$ is the minimization parameter. In order to study the properties of the superconductivity we have to minimize the above expression of $\lambda$ (\ref{lambda-equ}) with respect to $c$ and obtain $\lambda_{min}$.

We would like to mention a crucial point here: for the usual planar ($K = 0$) AdS scenario without noncommutativity (i.e. for $\theta = 0$), we have $f(r) = - \frac{2 M G}{r} + \frac{r^2}{L^2}$.
The horizon radius $r_+$ can be calculated from $f(r_+)=0$, which gives $r_+^3 = 2 M G $ (by considering the conventional choice $L=1$ for AdS-CFT correspondence). Thus one can obtain the expression of $\lambda^2$ using (\ref{eqn-lam-1}) (in probe limit) as
\beq
\lambda^2=\frac{\int_0^1\left[\left(1-z^3\right)(-2cz)^2+z(1-c~z^2)^2\right]dz}{\int_0^1\frac{(1-z)}{1+z+z^2}(1-c~z^2)^2dz}.
\label{lamusu} \eeq
Note that $M$ is canceled out from the numerator and the
denominator of (\ref{lamusu}) and $r_{+}$ does not appear explicitly. Thus the expression (\ref{lamusu}) is independent of the choice of $M$, and as well as of $r_+$ since $r_+^3 = 2 M G $. However, in this NC AdS scenario, the temporal component of metric tensor $g_{tt}$ is given by (\ref{metric-3}), using which the horizon radius is calculated from the relation $f_1(r_+) = 0$ (for $K=0,~L=1$) as follows:
\beq r_+^3 = 2 M G \left(1 - \frac{r_+}{\sqrt{\pi\theta}}e^{-\frac{r_+^2}{4\theta}}
\right). \label{rncads} \eeq
It is clear from the r.h.s. of (\ref{rncads}), that $r_{+}$ depends on $M$ as well as on $\theta$. Using the form $f(z)=-\frac{2 M G }{r_{+}}z +\frac{r_{+}^2}{z^2}$ and the relations (\ref{tqp}, \ref{lambda-equ}) the expression for the eigenvalue $\lambda$ turns out to be
\beq
\lambda^2=\frac{\int_0^1\left[\left(1-z^3-\frac{r_{+}}{\sqrt{\pi\theta}}e^{
		-\frac{r_{+}^2}{4\theta}}
	+z^2 \frac{ r_{+}}{\sqrt{\pi\theta}}e^{-\frac{r_{+}^2}{4\theta z^2}}\right)(-2cz)^2+
	z\left(1-\frac{r_{+}^3}{2\theta
		z^3}\frac{1}{\sqrt{\pi\theta}}e^{-\frac{r_{+}^2}{4\theta
			z^2}}\right)(1-c~z^2)^2\right]dz}
{\int_0^1\frac{1}{1+z+z^2}\left[1-z+\frac{2z^3-1}{1+z+z^2}\frac{r_{+}}{\sqrt{\pi\theta}}e^{-\frac{r_{+
			}^2}{4\theta}}-\frac{z^2}{1+z+z^2}
	\frac{r_{+}}{\sqrt{\pi\theta}}e^{-\frac{r_{+}^2}{4\theta
			z^2}}\right](1-c~z^2)^2dz}. \label{lamr+}
\eeq
Clearly in this case, the eigenvalue $\lambda$ not only depends on $M$ and $r_{+}$, but also on $\theta$. Subsequent results, such as $\zeta $ (the coefficient of $\sqrt{\rho}$ in the relation between the critical temperature $T_c$ and the charge density $\rho$) then becomes $M$-dependent as well as $\theta$-dependent. This is a highly non-trivial feature of the NC AdS model considered here. We speculate that it hints at a generalized form of AdS-CFT duality where the holographic superconductor may have other parameters, besides the charge density and chemical potential, as generally associated with it.\\

In the present noncommutative scenario the Hawking temperature $T_H$ \cite{hawking} is related to the horizon radius $r_{+}$ by the relation
\begin{equation}
T_H=\frac{3 r_+}{4\pi L^2} \left(1 + \frac{r_+}{3 \sqrt{\pi\theta}} e^{-\frac{r_+^2}{4 \theta}}   - \frac{ r_+^3}{ 6 \sqrt{\pi} \theta^{3/2}} e^{-\frac{r_+^2}{4 \theta}} \right). \label{T-r-rel}
\end{equation}
Substituting $r_{+c} = \sqrt{\frac{\rho}{\lambda}}$ from (\ref{fisol4}) in to the above expression we have the relation between the critical temparature and charged density as
\begin{equation}
T_c=\frac{3}{4\pi L^2} \sqrt{\frac{\rho}{\lambda_{min}}} \left(1 + \frac{1}{3 \sqrt{\pi\theta}}\sqrt{\frac{\rho}{\lambda_{min}}} e^{-\frac{\rho}{4 \theta \lambda_{min}}} - \frac{1}{ 6~ \theta^{\frac{3}{2}} \sqrt{\pi} } \left(\frac{\rho}{\lambda_{min}}\right)^{3/2} e^{-\frac{\rho}{4 \theta \lambda_{min}}} \right). \label{rel-Tc-rho}
\end{equation}
The above relation constitutes one of our principal results. From (\ref{rel-Tc-rho}) we see that $T_c$ not only depends on $\lambda_{min}$ and $\rho$, but also on noncommutative parameter $\theta$.
\begin{figure}[htb]
{\centerline{\includegraphics[width=8cm, height=6cm] {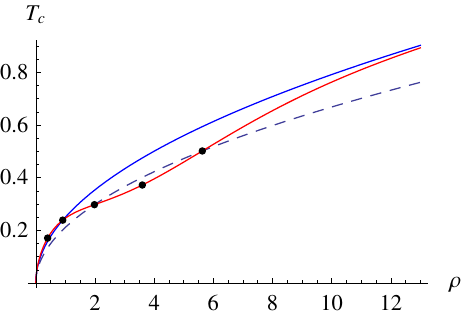}}}
\caption{{\it{Here the dashed curve corresponds to the usual relation between critical temparature $T_c$ and charged density $\rho$ where $T_c=0.225 \sqrt{\rho}$. The red and the blue curves correspond to NC relation (\ref{rel-Tc-rho}) and its linear order contribution (\ref{Tc-linear}) respectively. In this NC scenario we  got the minimum value of $\lambda_{min}$ $\approx$ 0.83. From this we get $\zeta=0.25$. The blue and red curves are plotted by taking this values of $\lambda_{min}$ and $\zeta$ for $\theta=0.5$. }}} \label{fig1}
\end{figure}
The linear order part (the part containing first order of $\sqrt{\rho}$) of (\ref{rel-Tc-rho}) is given by
\begin{equation}
T_c=\frac{3}{4\pi L^2} \sqrt{\frac{\rho}{\lambda_{min}}} \equiv \zeta \sqrt{\rho}~~~~~~~~\zeta=\frac{3}{4 \pi \lambda_{min} L^2}. \label{Tc-linear}
\end{equation}
In Figure 1 we have plotted (the red one) the NC corrected critical temperature $T_c$ against the charged density $\rho$. We have taken $\theta=0.5$ for the graph but we have checked that the structure of the curve remains same for any choice of $\theta$. One can see from this plot that for small values of charged density $\rho$ the red NC curve (\ref{rel-Tc-rho}) has some fluctuating behavior around its linear order part (\ref{Tc-linear}). For $\rho \leq 0.91$ the functional value of NC $T_c$ is greater than its linear part, but thereafter becomes lower and remains so. In fact, its value becomes less than compared to the normal one ( in picture this is dashed one for which $T_c = 0.225 \sqrt{\rho}$) in the range $1.98 \leq \rho \leq 5.63$. Thereafter its value continuously increases with $\rho$ and comes closer to its linear part ($\sqrt{\rho}$). However, since our intention is to study the superconductivity, so we are interested only in the increasing behavior of the critical temperature above normal case. Therefore, for the sake of brevity of calculations, here we concentrate on (\ref{Tc-linear}).
From (\ref{lamr+}), one can now determine numerical value of $\lambda_{min}$ and analyze (\ref{Tc-linear}), as studied in \cite{hart, dibakar}. In Tables 1, 2, 3 and 4 we have given some numerical estimate of $\zeta$ corresponding to the different values of $\lambda_{min}$, which further depends on the horizon radius $r_+$ and NC parameter $\theta$.

\section{Critical exponent and condensation value}
In this section, we are going to construct the condensation value of the condensation operator $J$ near the critical temperature $T=T_c$. In order to do that, we substitute (\ref{siassumption}) in (\ref{fi-eqn-2}) which gives
\beq
\phi''=\frac{\langle J\rangle^2}{r_{+}^2}B(z)\phi(z) \label{fidiff3} \eeq
where
$B(z)=\frac{r_{+}^2F^2(z)}{z^2f(z)}
-\frac{r_{+}^2F^2(z)}{z^2f^2(z)} \frac{2M G}{\sqrt{\pi\theta}}e^{-\frac{r_{+}^2}{4\theta z^2}} $.\\
At $T=T_c$, we know that $\psi=0$ and $\phi=\lambda r_{+c}(1-z)$ is the solution of $\phi''(z) = 0$.
Therefore, for a temperature $T$ close to $T_c$ we can consider the solution of (\ref{fidiff3}) to be of the form
\beq
\frac{\phi}{r_{+}}=\lambda(1-z)+\frac{\langle J\rangle^2}{r_{+}^2}\chi(z), \label{fiappro3}
\eeq
where $\frac{\langle J\rangle^2}{r_{+}^2}$ is a small parameter and $\chi(z)$ satisfies the boundary condition $\chi'(1)=\chi(1)=0$. Substituting the above relation (\ref{fiappro3}) in (\ref{fidiff3}) we get the differential equation for $\chi$ as,
\begin{equation}
\chi''(z)=\lambda r_+^2 (1-z)\frac{F^2(z)}{z^2 f(z)}\left(1-\frac{1}{f(z)}\frac{2M G}{\sqrt{\pi\theta}}e^{-\frac{r_{+}^2}{4\theta z^2}} \right) \label{chi-diff}
\end{equation}
Again, since (\ref{phiasyem1}) represents the asymptotic behavior of $\phi$, thus near the asymptotic boundary we can write
\beq \frac{\mu}{r_{+}}-\frac{\rho}{r_{+}^2}z=\lambda(1-z)+\frac{\langle J\rangle^2}{r_{+}^2}\chi(z)
=\lambda(1-z)+\frac{\langle
	J\rangle^2}{r_{+}^2}(\chi(0)+z\chi'(0)+...), \label{ficompare} \eeq
where we have expanded $\chi(z)$ about $z=0$. Comparing the coefficients of $z$ from both sides we have
\beq -\frac{\rho}{r_{+}^2}=-\lambda+\frac{\langle J\rangle^2}{r_{+}^2}\chi'(0).
\label{cirelation}\eeq
Now integrating (\ref{chi-diff}) between 0 to 1 and using the boundary condition $\chi'(1)=0$, we have the expression for $\chi'(0)$ as
\begin{equation}
\chi'(0) = -\lambda \int_0^1 \frac{(1-c z^2)^2}{(1 + z + z^2)} \left(1-\frac{r_+}{\sqrt{\pi\theta}}\frac{ z^2 (z e^{-\frac{r_{+}^2}{4\theta}} - e^{-\frac{r_{+}^2}{4\theta z^2}} )}{1 - z^3}
\right) dz \equiv -\lambda {\cal A}. \label{kappadiff5}
\end{equation}
Substituting $\chi'(0)$ from (\ref{kappadiff5}) in (\ref{cirelation}) and
using the relations $\lambda=\frac{\rho}{r_{+c}^2}$ and $T_c=\frac{3 r_{+c}}{4 \pi}$ (considering linear order term (\ref{Tc-linear})) we finally obtain the expression for condensation operator $J$ for $T\rightarrow T_c$ as
\beq
\langle J\rangle=\gamma T_c\sqrt{1-\frac{T}{T_c}} \label{Jvalue}
\eeq
where $\gamma=\frac{4\sqrt{2}\pi}{3\sqrt{{\cal A}}}.$
This relation (\ref{Jvalue}) is crucial for further study of other properties of the
noncommutative holographic superconductor.
Numerical estimates of $\gamma $ are provided below in Tables 1, 2, 3 and 4.\\

Note that the governing equations (\ref{siappro}) and (\ref{fiappro1}) of the scaler potential $\psi$ and the scaler field $\phi$ are obtained by considering the first order terms of O($\frac{2GM}{\sqrt{\pi\theta}}e^{-\frac{r^2}{4\theta}}$). If we expand $\frac{1}{f_1(r)}$ up to second order of O($\frac{2GM}{\sqrt{\pi\theta}}e^{-\frac{r^2}{4\theta}}$), then there will be some additional terms in (\ref{siappro}) and (\ref{fiappro1}). Though these terms do not change the asymptotic behavior of $\phi$, $\psi$ and equation (\ref{fisol4}), but affect the minimum value of $\lambda$ and $\chi'(0)$ through
(\ref{lambda-equ}) and (\ref{kappadiff5}) respectively. Thus the values of $\zeta$ and $\gamma$ will be affected due to the second order contributions. Considering the values of $\theta$ and $r_{+}$ from the tables one can obtain the maximum value of $(\frac{2GM}{\sqrt{\pi\theta}}e^{-\frac{r^2}{4\theta}})^2\approx O(10^{-2})$. Since the numerical values of $\zeta,~\gamma$ are approximately between $0.22-0.25$ and $8.2-7.6$ respectively, the O($10^{-2}$) corrections will affect these numerical values very slightly, thus for all practical purposes are negligible.

\section{Summary and Discussion}
In this paper, we have considered a NC charged AdS black hole and a scalar field coupled to gravity, thus introducing a hairy black hole. First we have studied the asymptotic behavior of the gauge and scalar field and explicitly show that there are no effects of noncommutativity on the physical parameters like charge density and chemical potential. Then we proceed to analyze the modified relation between critical temperature and charge density. Moreover, we have studied modified expressions for critical exponents and condensation values in this noncommutative context.

We have provided some numerical estimates in Table 1, 2, 3, 4. We consider the established upper bound of $\theta $ to be $\theta \leq (10~TeV)^{-2}$ \cite{jac}. In \cite{nicolini} the black hole mass $M$ is related to $\theta $ by $M \approx \sqrt{\theta}/G $ where the Newton's constant $G$ has been reinstated. This yields $M\approx 10^{33}$GeV, which, however is far below the mass of the astrophysical black holes. In Table 1 we have taken different values of $\theta$ to be lower than the above bound \cite{jac} and the corresponding black hole masses to be considerably larger than that in \cite{nicolini}. Expectedly for very large black hole and for different values of $\theta$ the relation $T_c=0.225\sqrt{\rho}$ \cite{hart} is recovered.

However, from Tables 2, 3 and 4 it is clear that for the same values of $\theta$, the critical temperature rises above the normal ($\theta=0$) case for different black hole masses (being chosen nearby the mass of the black hole considered in \cite{nicolini}). Interestingly, from Table 2, we find that for $\theta =0.5$ and $M=2.75~\sqrt{\theta}/G$ (which is close to the mass $M=2.4~\sqrt{\theta}/G$ considered in \cite{nicolini}) the value of $\zeta$ turns out to be $\zeta =0.25$, which is appreciably larger than the $\theta =0$ result ($\zeta =0.225$ \cite{hart}), indicating a larger critical temperature. Furthermore, one of the interesting finding from the Tables 2, 3, 4 is that, the critical temperature rise above the value $0.225$ if the mass of the Black hole $M$ lies within the interval $x\in[0.4,80]$, where $x=\frac{G M}{\theta^{3/2}}~Length^{-2}$. Moreover, if the Black hole mass is such that $x$ is around $x=55$ then the critical temperature is maximum since $\zeta\approx0.25$. And For larger $M$, i.e if $x>80$ then Table 1 shows that $\zeta$ stabilizes to $0.225$.

It will be interesting to consider non-zero vector potential in the noncommutative framework to study conductivity and other properties of the holographic superconductors. Furthermore, in the noncommutative extension considered here, we have confined ourselves in the probe limit (to neglect the backreaction of charged hair into the metric) to study the properties of holographic superconductor. It will be highly interesting to study these properties without considering the probe limit, which is our next goal. \\

{\bf Acknowledgement:} We thank Dibakar Roychowdhury for helpful discussions.

\vspace{0.5cm}
\begin{table}[htb]\small\addtolength{\tabcolsep}{-0.4pt}
	\centering
	\small\begin{tabular}{|c|c|c|c|c|c|c|c|c|}
		\hline { \bf $\theta$}  &{\bf $M$ }&{\bf $r_{+}$ }&{\bf $c$ }&{\bf $\lambda^2$ }
		&{\bf $\zeta$}&{\bf $\gamma$}\\
		
		\hline 0.5  & $ 10^3\sqrt{\theta}/G$ & 11.22462 & 0.238901 & 1.26832 & 0.225 & 8.07\\
		\hline 0.1  & $10^3\sqrt{\theta}/G$ & 3.984220 & 0.238901 & 1.26832 & 0.225 & 8.07\\
		\hline 0.01  & $10^3\sqrt{\theta}/G$ & 5.848036 & 0.238901 & 1.26832 & 0.225 & 8.074\\
		
		\hline
	\end{tabular}
	\caption{\it Here we have considered $G=1$. From the above table one can see that for a large mass black holes, the relation between the critical temperature and charged density $(T_c=\zeta \sqrt{\rho})$ remains unchanged for different values of $\theta$.}
	\label{tab1}
\end{table}
\newpage
\begin{table}[htb]\small\addtolength{\tabcolsep}{-0.4pt}
	\centering
	\small\begin{tabular}{|c|c|c|c|c|c|c|c|c|}
		\hline { \bf $\theta$}  &{\bf $M$ }&{\bf $r_{+}$ }&{\bf $c$ }&{\bf $\lambda^2$ }
		&{\bf $\zeta$}&{\bf $\gamma$}\\
		
		\hline 0.5  & $40\sqrt{\theta}/G$ & 3.836269 & 0.23811 & 1.26683 & 0.225 & 8.07\\
		\hline 0.5  & $5\sqrt{\theta}/G$ & 1.683034 & 0.132345 & 1.04226 & 0.236 & 7.71\\
		\hline 0.5  & $2.75\sqrt{\theta}/G$ & 1.290122 & 0.082962 & 0.827768 & 0.250 & 7.63\\
		\hline 0.5  & $0.5\sqrt{\theta}/G$ & 0.7314238 & 0.245897 & 1.12218 & 0.232 & 8.1\\
		\hline 0.5  & $0.2\sqrt{\theta}/G$ & 0.559243 & 0.2757998 & 1.265862 & 0.225 & 8.2\\
		\hline
	\end{tabular}
	\caption{\it Here we have considered $G=1$. From the above table one can see that for a fixed value of  $\theta=0.5$, as $M$ decreases the critical temperature first rises (as $T_c=\zeta \sqrt{\rho}$) above the normal value 0.225 and reaches to maximum value ($0.25\sqrt{\rho}$) and then decreases.}
	\label{tab2}
\end{table}

\begin{table}[htb]\small\addtolength{\tabcolsep}{-0.4pt}
	\centering
	\small\begin{tabular}{|c|c|c|c|c|c|c|c|c|}
		\hline { \bf $\theta$}  &{\bf $M$ }&{\bf $r_{+}$ }&{\bf $c$ }&{\bf $\lambda^2$ }
		&{\bf $\zeta$}&{\bf $\gamma$}\\
		
		\hline 0.1  & $8\sqrt{\theta}/G$  & 1.715632 & 0.23811 & 1.26683 & 0.225 & 8.07\\
		\hline 0.1  & $\sqrt{\theta}/G$ & 0.752676 & 0.0927152 & 0.943411 & 0.242 & 7.62\\
		\hline 0.1  & $0.55\sqrt{\theta}/G$ & 0.576960 & 0.082962 & 0.827768 & 0.250 & 7.63\\
		\hline 0.1  & $0.1\sqrt{\theta}/G$ & 0.327102 & 0.245898 & 1.12218 & 0.232 & 8.1\\
		\hline 0.1  & $0.04\sqrt{\theta}/G$ & 0.250101 & 0.275800 & 1.265862 & 0.225 & 8.2\\		
		\hline
	\end{tabular}
	\caption{\it Here we have considered $G=1$. From the above table one can see that for a fixed value of $\theta=0.1$, as $M$ decreases the critical temperature first rises (as $T_c=\zeta \sqrt{\rho}$) above the normal value 0.225 and reaches to maximum value ($0.25\sqrt{\rho}$) and then decreases.}
	\label{tab2}
\end{table}
\begin{table}[htb]\small\addtolength{\tabcolsep}{-0.4pt}
	\centering
	\small\begin{tabular}{|c|c|c|c|c|c|c|c|c|}
		\hline { \bf $\theta$}  &{\bf $M$ }&{\bf $r_{+}$ }&{\bf $c$ }&{\bf $\lambda^2$ }
		&{\bf $\zeta$}&{\bf $\gamma$}\\
		
		\hline 0.01  & $0.8\sqrt{\theta}/G$  & 0.542530 & 0.23811 & 1.26683 & 0.225 & 8.07\\
		\hline 0.01  & $0.1\sqrt{\theta}/G$ & 0.238017 & 0.0927151 & 0.94341 & 0.242 & 7.62\\
		\hline 0.01  & $0.055\sqrt{\theta}/G$ & 0.182451 & 0.082462 & 0.827768 & 0.250 & 7.63\\
		\hline 0.01  & $0.01\sqrt{\theta}/G$ & 0.103439 & 0.245897 & 1.12218 & 0.232 & 8.10\\
		\hline 0.01  & $0.004\sqrt{\theta}/G$ & 0.079089 & 0.275799 & 1.265862 & 0.225 & 8.2\\
		\hline
	\end{tabular}
	\caption{\it Here we have considered $G=1$. From the above table one can see that for a fixed value of $\theta=0.01$, as $M$ decreases the critical temperature first rises (as $T_c=\zeta \sqrt{\rho}$) above the normal value 0.225 and reaches to maximum value ($0.25\sqrt{\rho}$) and then decreases.}
	\label{tab3}
\end{table}

\ed

We find that larger values of $\theta$ tend to lower the critical temperature of the holographic superconductor. This is our principal result.

More specifically, the model describing  phase transitions in high $T_c$ superconductors consists of a charged scalar field minimally coupled to an Abelian gauge field in an AdS Black hole background \cite{holo}. The Black hole admits scalar hair at a temperature $T$ below a certain critical temperature $T_c$ by the mechanism of breaking of a local $U(1)$ symmetry near the event horizon of the Black hole \cite{gubser}. The emergence of a hairy AdS Black hole implies the
formation of a charged scalar condensate in the dual CFTs by the AdS/CFT correspondence. The minimal model has been generalized in different directions in the AdS sector to study its effects on the CFT. An interesting extension was discussed in \cite{dibakar} that considered non-linear Born-Infield model instead of the usual Maxwell lagrangian. The analysis showed that the critical temperature of the Holographic Superconductor decreased with increase of the non-linear coupling. In the present article we study a completely different type of generalization of the gravity-field theory sector - Non-Commutative (NC) extension of the space-time. Our results indicate that NC effects are small but can lower the critical temperature of the Holographic Superconductor. It also turns out that the ratio of the NC parameter $\theta$ to the Black hole mass remains within a certain fixed value. Our results smoothly reduce to the conventional results for $\theta=0$.

We now study some thermodynamic properties for this NC AdS Reissner-Nordstrom type black hole (\ref{metric}). The Hawking temperature $T_H$ for a black hole is defined as \cite{hawking}
\beq
T_H = \frac{1}{4 \pi} \pa _r g_{00} |_{r = r_+}, \label{hawking}
\eeq
where $r_+$ is the outer black hole event horizon.
With this metric (\ref{metric}) in our hand, we go on to calculate the Hawking temperature $T_H$ for this black hole. However, due to the presence of the incomplete Gamma functions in the metric the calculations become very cumbersome and are not very illuminating either. For this reason, we study the case for large black holes (i.e. black holes with $r_+/\sqrt{\theta} >> 1$) for which the above metric takes the following form (using the asymptotic properties of incomplete gamma functions)
\beq
ds^2=-f_1(r)dt^2+\frac{dr^2}{f_1(r)}+r^2(dx^2+dy^2); $$$$
f_1(r) = K - \frac{2 M G}{r} + \frac{Q^2}{r^2} + \frac{r^2}{L^2} + \frac{2 M G}{\sqrt{\pi \theta}} e^{-\frac{r^2}{4 \theta}} - \frac{Q^2 G}{\sqrt{2} \pi \theta} e^{-\frac{r^2}{4 \theta}}, \label{metric1} \eeq